\documentclass[preprint,11pt]{elsarticle}

\usepackage{siunitx}
\usepackage{commath}
\usepackage{graphicx}
\usepackage{subcaption}
\usepackage{eurosym}
\usepackage{amsmath,amssymb,setspace,hyperref,color,enumitem,multirow,etoolbox}
\usepackage{cases}
\usepackage[version=3]{mhchem}
\usepackage{soul}
\usepackage[nomarkers,nolists,heads]{endfloat}
\usepackage{nomencl}
\makenomenclature

\usepackage{etoolbox}
\renewcommand\nomgroup[1]{%
  \item[\bfseries
  \ifstrequal{#1}{A}{Latin letters}{%
  \ifstrequal{#1}{B}{Greek letters}{%
  \ifstrequal{#1}{C}{Subscripts}{%
  \ifstrequal{#1}{D}{Superscripts}{}}}}%
]}

\definecolor{Gray}{gray}{0.5}

\journal{arXiv}
\hypersetup{%
            colorlinks=true, 
}
\begin{document}
  \hypersetup{%
              linktoc=all,  
              linkcolor=black, 
              urlcolor=black,
              citecolor=black
  }

\begin{frontmatter}
  \title{Optical characterisation of ilmenite by reflectance spectroscopy}
  \author[add1]{Jingjing Chen}
 \ead{jchen022@hku.hk}
  \author[add1]{Xiaobo Yin}
  \ead{xbyin@hku.hk}
\address[add1]{Department of Mechanical Engineering,
               The University of Hong Kong, Hong Kong, China}
\cortext[cor1]{}
\begin{abstract}
Bi-directional reflectance spectroscopy based on multiple scattering 
of particulate surfaces is employed in identifying the optical 
properties of ilmenite from laboratory reflectance measurements. 
However, the approach suffers from issues including: 
i) both $n$ and $k$ are to be extracted from a single spectroscopy spectrum, 
ii) imposing constraints of the Kramers--Kronig correlation relating 
spectral $n$ and $k$ is weakened by its fundamental insensitivities, and
iii) incapability in addressing intrinsic strong absorption 
features of absorbing materials.
We resolve these issues by employing additional 
optical information of directional--hemispherical reflectance
and emissivity/absorptivity for a slab and a stratified 
multi-layer medium of the material, respectively.
The accompanied analyses consist of radiative transfer in a slab 
medium investigated using the two-flux approximation method and 
electromagnetic radiation propagation in a stratified multi-layer 
medium investigated using the electric filed transfer matrix.
We further find that an understanding of the internal scattering 
coefficient of grains in the multiple scattering model paves the
way in successfully predicting absorption features of materials.
A wavelength-dependent internal scattering coefficient of 
ilmenite is then found to be 50 \si{\per\micro\metre} and 
$10^{-7}$ \si{\per\micro\metre} in regions of strong absorption
($<$\SI{7}{\micro\metre}, $>$\SI{13}{\micro\metre}) 
and high transmission (7--\SI{13}{\micro\metre}), respectively. 
The value of the refractive index $n$ varies weakly on the wavelength.
A pronounced change in the determined absorptive index $k$ 
with the wavelength is observed.
Low values of the absorptive index $k$ on the magnitude of $10^{-2}$ 
are obtained in the transmission window spectral range.
In the strong absorption spectral range starting from \SI{13}{\micro\metre}, 
values of the absorptive index $k$ are higher than 0.1.
\end{abstract}
\begin{keyword}
Radiative transfer\sep
Multiple scattering \sep
Reflectance spectroscopy \sep
Two-flux approximation \sep
$N$-phase multi-layer  
\end{keyword}
\end{frontmatter}
\setlength\parskip{8pt}
\onehalfspacing
\normalsize
\clearpage
\newpage
\section{Introduction}
Detailed optical properties (refractive index $n$ and absorptive
index $k$) of materials are necessary for forward modelling of 
spectra which is to calculate the spectrum of a designed phase 
mixture as employed in developing photonic materials 
\cite{l-009-197,np-016-268,olt-142-107265,nature-515-540,science-350-1062}  
and inverse modelling which is to extract the properties from the 
spectrum of a remotely observed surface as encountered in reflectance 
spectroscopy \cite{rs-016-268,rs-008-1807,jgr-089-6329,op-045-3427}.
The frequently used method for determining optical properties of 
materials is depositing a thin film on a substrate by sputtering
techniques and then identifying the properties using ellipsometry 
by measuring the change in the polarization state as light reflects 
from the thin film surface \cite{tompkins-uge-1993}. 
However, materials such as ilmenite cannot be evaporated 
and deposited in a uniform coating, thus ruling out ellipsometry
as a determination method \cite{icarus-361-114331}.
An alternative approach, which is a combination of measurements 
of bi-directional reflectance of particulate media and subsequent 
analysis of radiative transfer within using the multiple 
scattering theory, was proposed by Hapke \cite{jgr-086-3039}.
Considering both $n$ and $k$ are to be determined from a single 
reflectance spectrum, currently, the standard approach is employing 
a constant $n$, which is taken from existing reference data of 
the material, and iteratively varying $k$ to make the calculated 
and measured reflectance spectrum to agree with each other. 
Roush \cite{jgr-112-e100} later improved this approach by subsequently 
using a subtractive Kramers--Kronig analysis to allow determination of 
$n$ as a function of wavelength. 
However, this handling of $n$ requires a prior understanding of 
investigated materials, and suffers from fundamental insensitivity to 
spectral features and changes in $k$ across the entire spectral range.
The approach has been applied to determine the optical properties
of transparent \cite{jgr-103-1703,jgr-095-14743,icarus-086-355,jgr-099-10867,jgr-112-e100} and opaque materials \cite{icarus-361-114331}.
However, it is recognized that the approach provides a poor determination
of the optical properties in infrared regions where absorbing materials 
exhibit strong absorption specifications \cite{jgr-112-e100,icarus-361-114331}. 

In this work, we aim to resolve these issues encountered
in the standard reflectance spectroscopy approach based on 
the Hapke radiative transfer model and identify the optical
properties of dark, opaque ilmenite.
For solving the issue of two unknowns $n$, $k$ to be determined from 
a single reflectance spectrum for powdered materials, additional 
information of radiative transfer in a slab medium is introduced. 
Specifically, the determination of the refractive index $n$ is achieved 
by a distinct reflectance spectroscopy for the pellet material.
By considering the pellet as a plane-parallel slab medium, the radiative
transfer within is analyzed using the two-flux approximation method 
\cite{howell-trht-2020,modest-rht-2021,josaa-023-0091,jqsrt-272-107754},
and the directional--hemispherical reflectance is found to be 
a function of the refractive index $n$ of materials.
For solving the issue of incapable of identifying strong absorption 
characteristics in the standard Hapke model, a thorough investigation
on the radiative properties involved in the model is conducted. 
Specifically, the internal scattering coefficient parameter 
which is neglected in the standard approach turns out to be the key.
In addition, the validation of the investigation on the parameter 
and the combined reflectance spectroscopy is provided by 
scrutinizing the absorptivity/emissivity of the thin film media by
embedding the particulate material into transparent plastic matrix. 
The proposed approach is then employed to identify the optical 
properties of opaque ilmenite.
\section{Measurement Data}
Three types of samples made from ilmenite---particulate,
pellet, and film---are used for investigating the
optical properties of ilmenite in this study.
The bi-directional reflectance, directional--hemispherical 
reflectance, and absorptivity/emissivity are obtained for 
the particulate, pellet, and film media, respectively.
\subsection{Bi-directional reflectance of particulate media}
\begin{figure}[!hbt]
  \centering
  \includegraphics[width=8cm]{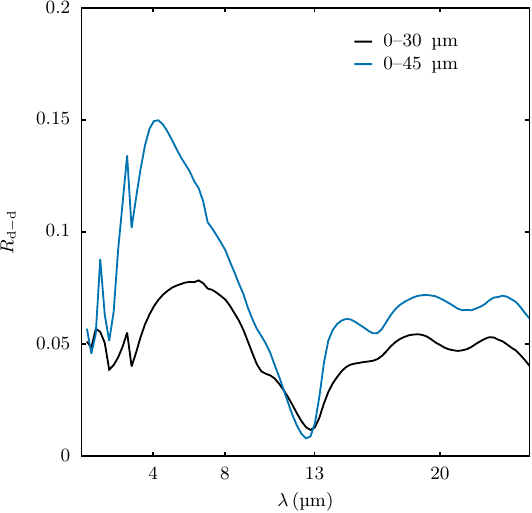}
  \caption{Bi-directional reflectance spectra of particulate ilmenite 
           from RELAB spectral library.}
  \label{fig-powder}
\end{figure}

The RELAB online spectral library archives spectral data and sample 
preparation, characterization, acquisition, and spectral measurements.
Assuming the composition is independent of grain size, additional 
leverage is achieved by using reflectance spectra of two different 
grain sizes of ilmenite.
The ilmenite powder samples investigated in this study 
are 0--\SI{30}{\micro\metre} (IL-M1O-005) and 
0--\SI{45}{\micro\metre} (IL-LPK-007).
Bi-directional reflectance spectra are for incidence and 
emission angles of \SI{30}{\degree} and \SI{0}{\degree}, respectively.
The bi-directional reflectance of particulate ilmenite with the two 
particle size distributions is shown in Fig.~\ref{fig-powder}.
More details on the measurement approach can be found in 
The RELAB spectral library and \cite{jgr-112-e100}.
\subsection{Directional--hemispherical reflectance of pellet media}
\begin{figure}[!hbt]
  \centering
  \includegraphics[width=8cm]{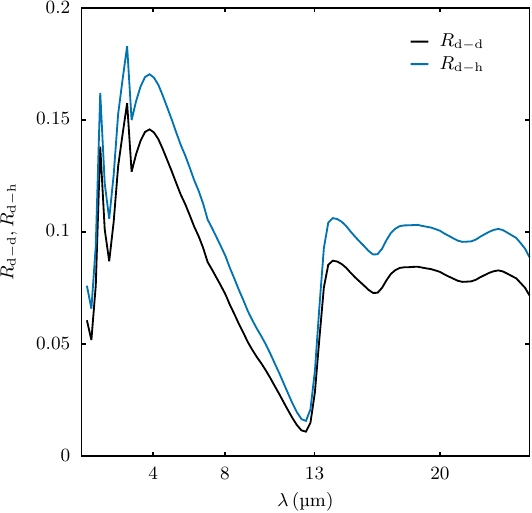}
  \caption{Bi-directional reflectance of pellet ilmenite from RELAB spectral 
           library and calculated directional--hemispherical reflectance 
           employing the Shkuratov and Grynko model \cite{icarus-173-16}.}
  \label{fig-pellet}
\end{figure}

Similarly, Bi-directional reflectance for pellet ilmenite (IL-LPK-007-P) 
is obtained from RELAB spectral library (see Fig.~\ref{fig-pellet}).
The directional--hemispherical reflectance is calculated 
using Shkuratov and Grynko relation 
$\log R_{\text{d}-\text{d}}=1.088 \log R_{\text{d}-\text{h}}$ 
\cite{icarus-173-16},
where $R_{\text{d}-\text{d}}$ and $R_{\text{d}-\text{h}}$ are 
bi-directional and directional--hemispherical reflectance, respectively.
The effect of material volume fraction on the relation is neglected
and detailed analysis is beyond the scope of the current paper.
The resulting data is also plotted in Fig.~\ref{fig-pellet}.
\subsection{Emissivity/absorptivity of film media}
\begin{figure}[!hbt]
  \centering
  \includegraphics[width=8cm]{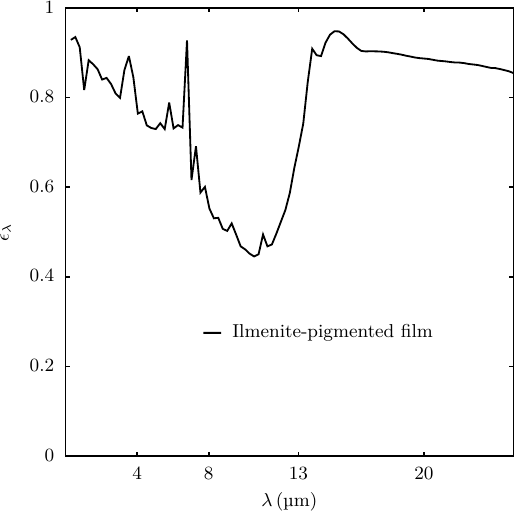}
  \caption{Emissivity/absorptivity of ilmenite-pigmented low-density
           polyethylene film backed with aluminium substrate.}
  \label{fig-film}
\end{figure}

The ilmenite and low-density polyethylene powders are mixed at 
a volume ratio of 1:9. The film material with a thickness of
\SI{50}{\micro\metre} on aluminium is fabricated by hot-pressing
the mixed powder at 160\si{\degreeCelsius} and 25 \si{\mega\pascal} 
onto a highly reflecting aluminium for 10 seconds. 
The spectral reflectance of the film fabricated in the solar spectrum 
(0.3--\SI{2.5}{\micro\metre}) and mid-infrared (2.5--\SI{25}{\micro\metre}) 
are characterized using UV-VIS-NIR spectrophotometer (Cary 5000, Agilent) 
equipped with an integrating sphere (DRA-2500, Agilent) and 
Fourier transform infrared spectrometer (Nicolet is50) with 
a gold-coated integrating sphere (PIKE technologies), respectively.
Employing the Kirchhoff's law, the emissivity/absorptivity is
obtained as $\epsilon=A=1-R$ and shown in
Fig.~\ref{fig-film}, where $\epsilon$, $A$, and $R$
are emissivity, absorptivity, and reflectance, respectively.

\section{Analytic methods}
The phenomena involved in determining and evaluating optical 
constants of materials in this work consist of light propagation 
and interaction in particulate, slab, and multi-layer media.
For the particulate media, an analysis of the radiative transfer  
within describing the scattering of light from 
particulate surfaces is performed \cite{jgr-086-3039}. 
The resulting analytic solution relates bi-directional reflectance 
with the single scattering albedo and the single scattering albedo 
with the optical constants.
For investigating radiative transfer in the plane-parallel slab of 
medium, the two-flux approximation is employed and the result of
directional--hemispherical reflectance is obtained as a function of
the refractive index of the material \cite{josaa-023-0091,jqsrt-272-107754}.
The emissivity/absorptivity data of the multi-layer film 
is scrutinized for revealing and evaluating the absorption 
features of the material in the infrared region.
Its underlying physics involved is light propagation in a multi-layer 
medium and electric fields induced by plane electromagnetic 
radiation within are derived \cite{josa-058-380}.
\subsection{Analysis of radiative transfer in particulate media 
by the multiple scattering theory}
\begin{figure}[!hbt]
  \centering
  \includegraphics[width=12cm]{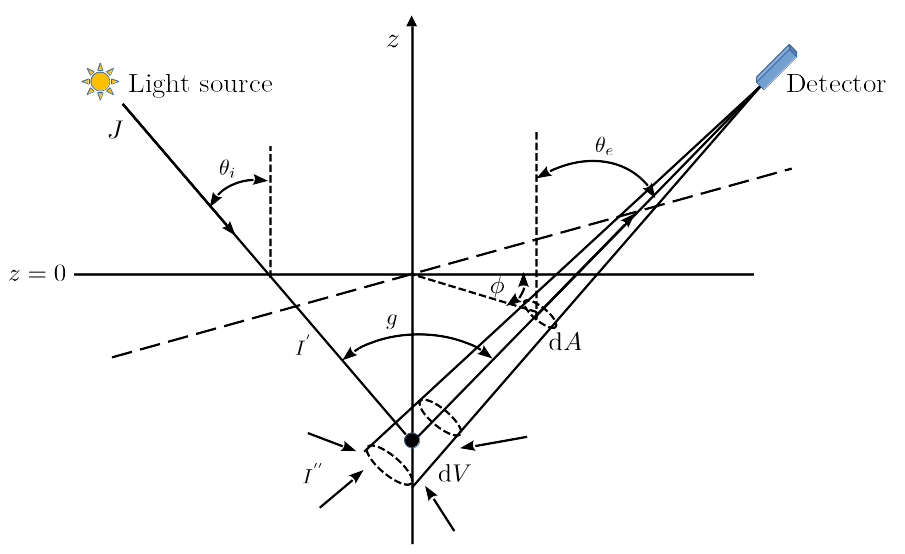}
  \caption{Schematic of the scattering of light from particulate 
           surfaces.}
  \label{fig-sch-hapke}
\end{figure}

For a plane surface containing absorbing particles, the medium 
is illuminated by collimated light traveling into direction with 
an incidence angle of $\pi-\theta_{\text{i}}$.
The medium is observed with an emission angle of $\theta_{\text{e}}$.
The description of light attenuation and scattering from individual 
particulate surface is schematically shown in Fig.~\ref{fig-sch-hapke}. 
Assuming that the particles are irregular and arranged with no
particular orientation, coherent effects average out.
Taking multiple scattering, mutual shadowing and opposition 
effects into account, the bi-directional reflectance is 
obtained as \cite{jgr-086-3039},
\begin{equation}
 R_{\text{d}-\text{d}} = \dfrac{\omega}{4 \pi} \dfrac{\mu_{0}}{\mu_{0}+\mu}
                         \bigl\{ \left[ 1+B(g)\right] \Phi(g) 
                         +H(\mu_{0}) H(\mu)-1 \bigr\},
\end{equation}
where $\mu_{0}=\cos \theta_{\text{i}}$, 
$\mu=\cos \theta_{\text{e}}$, and $g$ is phase angle.
$\omega$, $B$, $\Phi$, and $H$ are single scattering albedo,
backscattering function, scattering phase function,
and multiple-scattering function, respectively, and are written as,
\begin{align}
  \omega =& \dfrac{Q_{\text{S}}}{Q_{\text{E}}} ,\\
  B(g)   =& \exp (- \omega^{2}/2) \bigl\{ 1-\dfrac{\tan \mid g \mid }{2 h} 
            \left [3-\exp(-h/\tan \mid g \mid)\right] 
            \left[1-\exp(-h/\tan \mid g \mid)\right] \bigr \},\\
  \Phi(g)=& 1-0.4 \cos (g) +0.25\left[ 1.5 \cos^{2}(g) -0.5  \right ],\\
  H(\mu) =& \dfrac{1+2\mu}{1+2 \chi \mu}.
\end{align}
$Q_{\text{S}}$, and $Q_{\text{E}}$ are scattering efficiency and 
extinction efficiency, respectively, which can be calculated 
using the Mie theory \cite{howell-trht-2020,modest-rht-2021}. 
Hapke derived simplified analytical solutions, which are 
functions of internal scattering coefficient $s$ and internal 
absorption coefficient $\alpha=\dfrac{4 \pi k}{\lambda}$ \cite{jgr-086-3039}.
The relations are omitted here for brevity.
$\chi=(1-\omega)^{1/2}$, and $h=0.1$.

The process for determining the optical properties is described as follows.
Observing that the derived bi-directional reflectance is a function of 
the single scattering albedo, thus the single scattering albedo is 
extracted from the measured bi-directional reflectance.
Then, the single scattering albedo relates to 
the grain size and the optical properties of the material.
We assume a representative particle size corresponding to the 
average of the reported sieve size range for each powder samples: 
\SI{15}{\micro\metre} for for 0--\SI{30}{\micro\metre} separate and  
\SI{22.5}{\micro\metre} for for 0--\SI{45}{\micro\metre} separate.
The conversion of the single albedo to the optical 
constants thus removes the effects of particle size.
\subsection{Analysis of radiative transfer in slab media 
by the two-flux approximation}
\begin{figure}[!hbt]
  \centering
  \includegraphics[width=12cm]{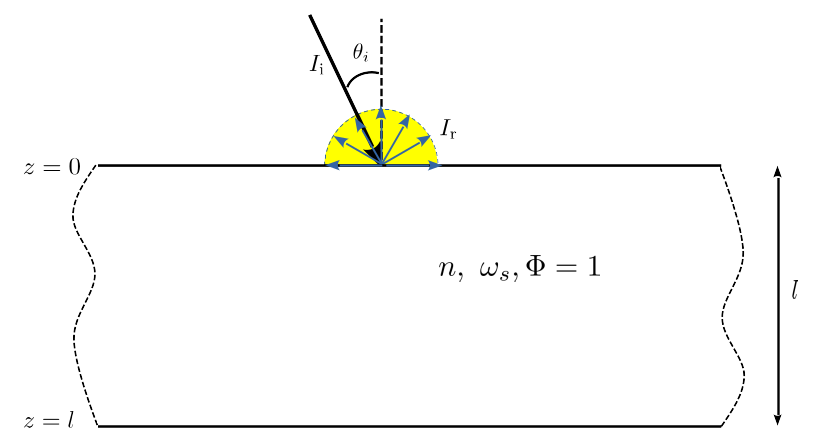}
  \caption{Schematic of a plane-parallel layer of an absorbing,
           scattering and refracting medium.}
  \label{fig-sch-tf}
\end{figure}

Consider the problem of light incident onto a pellet medium.
Treating the pellet as a one-dimensional plane-parallel layer of an
absorbing, refracting, and scattering medium, as shown in
Fig.~\ref{fig-sch-tf}, the two-flux approximation method is employed 
to analyze the radiative transfer in the medium.
The method is subject to the following assumptions:
(i) the investigated materials are homogeneous and isotropic, 
(ii) radiation is isotropic scattered,
(iii) the surface is isotropic and diffuse,
(iv) the sample is optically thick.
The directional--hemispherical reflectance is then 
obtained as \cite{josaa-023-0091,jqsrt-272-107754}: 
\begin{equation}
  R_{\text{d}-\text{h}} = R_{1}+(1-R_{1}) \frac{\gamma}{2n^{2}}
      \frac{\omega_{\text{s}}}{1-\omega_{\text{s}}}
      \frac{\kappa^{2}}{(1+\kappa)(\frac{2\kappa}{1+\eta}+\kappa)},
\end{equation}
where
\begin{align}
  \kappa^{2} &= \frac{4}{(1+\eta)^{2}}\frac{1-\omega_{\text{s}}}{1-
                \omega_{\text{s}}\eta},    \\
  \gamma     &= \frac{1-R_{1}}{1+R_{1}},        \\
  \eta       &= \sqrt{1-\frac{1}{n^{2}}}.
\end{align}
$\omega_{\text{s}} =\dfrac{s}{s+\alpha}$ is the scattering
albedo of the slab medium.
The interface reflectivity $R_{1}$ is obtained by averaging
the Fresnel reflectivity over the hemisphere,
\begin{equation}
  R_{1}=\frac{1}{2}+\frac{(3n+1)(n-1)}{6(n+1)^{2}}+
        \frac{n^{2}(n^{2}-1)^{2}}{(n^{2}+1)^{3}}\ln\left(
        \frac{n-1}{n+1}\right)-\frac{2n^{3}(n^{2}+2n-1)}{(n^{2}+1)(n^{4}-1)}+
        \frac{8n^{4}(n^{4}+1)}{(n^{2}+1)(n^{4}-1)^{2}}\ln n.
\end{equation}

It is observed that the directional--hemispherical reflectance is a
function of the refractive index $n$.
Thus, the refractive index $n$ is identified from fitting with
the measured reflectance spectra of the pellet material.
The value is then used in the previous inverse process based on the Hapke 
radiation model for identifying the absorptive index $k$.
The whole process containing these two steps is iterated until 
changes in $n$ and $k$ are insignificant. 
\subsection{Analysis of radiative transfer in multi-layer stratified media 
by the electric field transfer matrix}
\begin{figure}[!hbt]
  \centering
  \includegraphics[width=12cm]{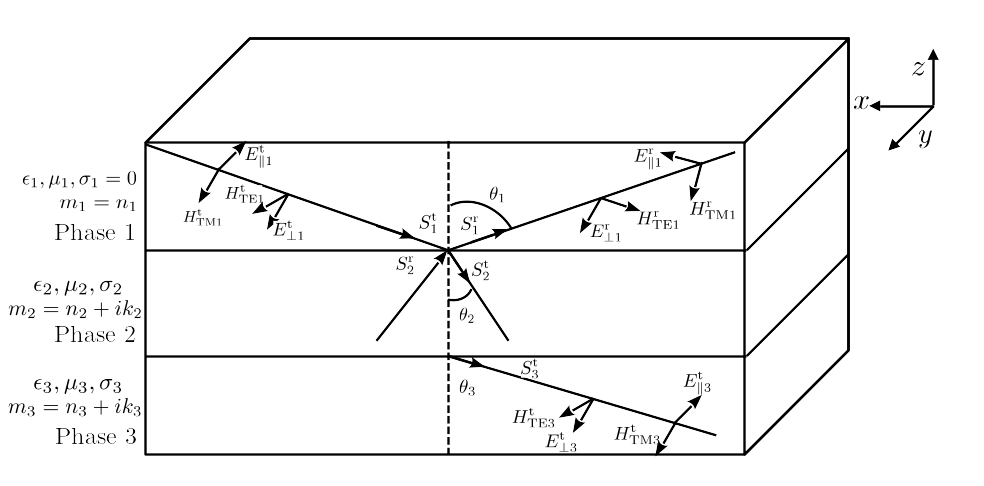}
  \caption{Plane coherent electromagnetic radiation interaction 
           in a three-phase medium.}
  \label{fig-sch-layer}
\end{figure}

When discussing reflection from a film layer on top of a distinct 
substrate, reflection and transmission of electromagnetic
radiation from one or more phases in the form of stratified
medium of multiple layers are investigated (see Fig.~\ref{fig-sch-layer}).
Specifically, the medium system of ilmenite-pigmented film deposited 
on aluminium substrate in this work is treated as a $N$-phase ($N=3$) 
stratified medium, which consists of air, ilmenite-pigmented 
low-density polyethylene film, and aluminium substrate. 
This results in $N-1$ surfaces of discontinuity at $z=z_{k}$
($k=1,2, ..., N-1$).
Each Phase ($z_{k-1}<z<z_{k}$) is characterised with its
individual thickness $h_{j}$, dielectric constant $\epsilon_{j}$, 
permeability $\mu_{j}$,
conductivity $\sigma_{j}$, complex index of refraction 
$m_{j}=n_{j}+ik_{j}$, and transfer matrix $M_{j}$.
The tangential fields at the first boundary, $z_{1}$, are related 
to those at the final boundary, $z_{N-1}$, by \cite{josa-058-380}
\begin{equation}\label{eq-matrix}
 \begin{bmatrix}
  U_{1} \\
  V_{1}
  \end{bmatrix}
  = M_{2}M_{3} \cdot \cdot \cdot M_{N-1}
 \begin{bmatrix}
  U_{N-1} \\
  V_{N-1}
  \end{bmatrix}
  = M
 \begin{bmatrix}
  U_{N-1} \\
  V_{N-1}
  \end{bmatrix}  ,
\end{equation}
where $U_{k}$ and $V_{k}$ are the tangential components of the 
electric (E)or magnetic (H) field amplitudes at the $k$-th boundary.
Specifically, for parallel ($\parallel$) polarization,
$U_{k}=H_{y}^{0}$, and $V_{k}=E_{x}^{0}$,
for perpendicular ($\perp$) polarization,
$U_{k}=E_{y}^{0}$, and $V_{k}=H_{x}^{0}$.
$M_{j}$ is the characteristic matrix for the $j$-th layer,
it is written as
\begin{equation}
 M_{j} =
 \begin{bmatrix}
 \cos \beta_{j}  &  \dfrac{-i}{p_{j}}\sin \beta_{j}  \\
 -ip_{j} \sin \beta_{j} &  \cos \beta_{j}
 \end{bmatrix}
,
\end{equation}
and
\begin{equation}
 M_{j} =
 \begin{bmatrix}
 \cos \beta_{j}  &  \dfrac{-i}{q_{j}}\sin \beta_{j}  \\
 -iq_{j} \sin \beta_{j} &  \cos \beta_{j}
 \end{bmatrix}
,
\end{equation}
for TE and TM polarization, respectively, where 
$\beta_{j}=\dfrac{2\pi}{\lambda} m_{j} \cos \theta_{j} h_{j}$, 
$p_{j}=(\epsilon_{j}/\mu_{j})^{1/2} \cos \theta_{j}$, and
$q_{j}=(\mu_{j}/ \epsilon_{j})^{1/2} \cos \theta_{j}$.

The reflectance is thus obtained as
\begin{align}
  R_{\parallel}&=\bigg | \dfrac{ H_{y1}^{0\text{r}}}{H_{y1}^{0\text{t}}}
                 \bigg |^{2}=\bigg | \dfrac{(M_{11}+M_{12}q_{N})q_{1} -
                 (M_{21}+M_{22}q_{N})}{(M_{11}+M_{12}p_{N})q_{1} 
                  +(M_{21}+M_{22}q_{N})  }  \bigg |^{2}   ,  \\
  R_{\perp}    &=\bigg | \dfrac{ E_{y1}^{0\text{r}}}{E_{y1}^{0\text{t}}}
                 \bigg |^{2} =\bigg | \dfrac{(M_{11}+M_{12}p_{N})p_{1} -
                 (M_{21}+M_{22}p_{N})}{(M_{11}+M_{12}p_{N})p_{1}   
                  +(M_{21}+M_{22}p_{N})  }\bigg |^{2}  ,
\end{align}
where $M_{ij}$ are the elements of the matrix $M$ in Eq.~\eqref{eq-matrix},
superscripts r and t represent reflected and transverse components.
Employing the optical properties of ilmenite determined from the 
previous subsections using the Hapke and two-flux approximation models,
the reflectance and thus emissivity/absorptivity of the fabricated 
ilmenite-pigmented low-density polyethylene film on aluminium
is obtained. 
\section{Results and discussions}
The bi-directional reflectance for the particulate media of the
$<$\SI{30}{\micro\metre} and $<$\SI{45}{\micro\metre} grain size 
fractions of ilmenite are used in the Hapke radiative transfer 
model to determine the absorptive index $k$. 
The refractive index $n$ is extracted from the directional--hemispherical
reflectance for the pellet medium of ilmenite by an inverse process 
based on the two-flux approximation radiative transfer model.
It is found that the internal scattering coefficient, $s$, in the Hapke
model plays an important role in the optical property identification. 
In the absence of a prior understanding of the material, the parameter 
is set to a small or even zero constant value across the entire spectral 
range in the standard approach \cite{jgr-112-e100,jgr-103-1703}.
Accurate determination of the optical constants has been provided for
materials showing weak absorption characteristics. 
However, inaccuracies in estimates of the absorptive
index are observed to be introduced in regions of strong absorption.
Thus, the determination of the appropriate value of the internal
scattering coefficient is required to accurately obtain the 
optical properties of absorbing materials.
\subsection{Constant internal scattering coefficient}
\begin{figure}[!hbt]
  \centering
  \includegraphics[width=8cm]{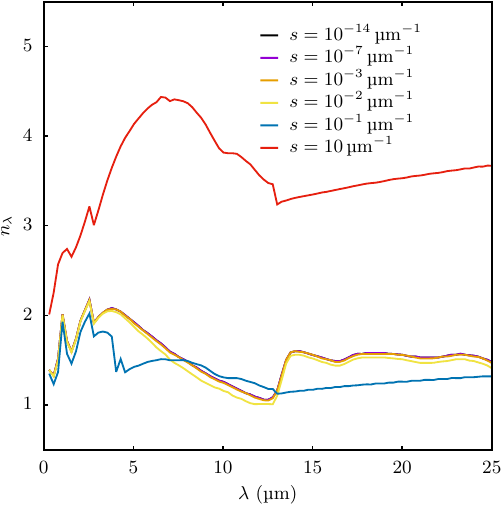}
  \caption{Comparison of the refractive index derived for varying constant 
           internal scattering coefficient.}
  \label{fig-con-s-n}
\end{figure}

\begin{figure}[!hbt]
  \centering
  \includegraphics[width=8cm]{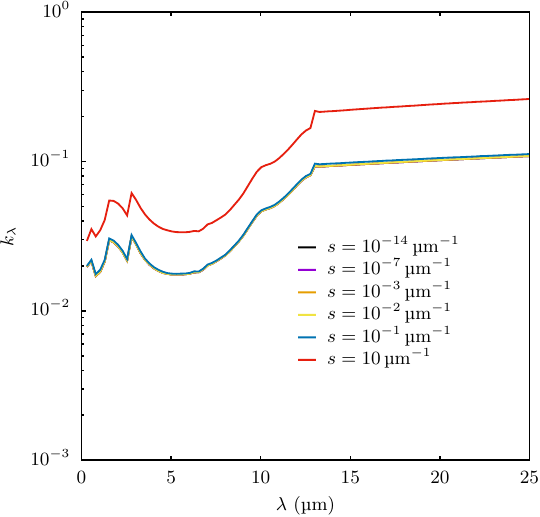}
  \caption{Comparison of the absorptive index derived for varying constant
           internal scattering coefficient.}
  \label{fig-con-s-k}
\end{figure}

First, assuming that the internal scattering coefficient $s$ is 
independent of the wavelength, the calculation procedure of
determining the optical constants is performed for employing the 
values of $s=10^{-14}, 10^{-7}, 10^{-3},
10^{-2}, 10^{-1}, 10$ (units of \si{\per\micro\metre}).
The refractive index is then determined and plotted 
in Fig.~\ref{fig-con-s-n}.
It is seen that values of the refractive index $n$ obtained show 
negligible difference for $s<10^{-2}\,\si{\per\micro\metre}$.
The determined absorptive index is also plotted in Fig.~\ref{fig-con-s-k}.
The values of the absorptive index $k$ align very well
when employing small values of $s<10^{-1}\,\si{\per\micro\metre}$.
In addition, reflectance spectra are reconstructed using optical
properties determined to examine the fitting performance.
Good agreement is achieved when employing these small internal 
scattering coefficient values. 
Then, further scrutiny is engaged by increasing the values of 
the internal scattering coefficient.
Specifically, with the internal scattering coefficient increasing to 
$s=10^{-1}$ \si{\per\micro\metre}, the values of $n$ deviate from 
those of employing smaller $s$.
When observing the corresponding phenomenon in the absorptive index,
it is found that considerably higher values are retrieved 
when $s$ is increased to 10 \si{\per\micro\metre}.
The calculated reflectance spectra also correspondingly start to 
show poor agreement with the measurement reflectance data.
Combing both of the analyses for $n$ and $k$, it is concluded 
that the internal scattering coefficient for ilmenite material is 
small enough in the range of $s<10^{-2}\,\si{\per\micro\metre}$, 
and become large in the range of $s>10^{-1}\,\si{\per\micro\metre}$. 
The accurate fitting of calculation with measurement reflectance data
across the entire spectral range requires a small enough value of 
the internal scattering coefficient, which is corroborated in 
relevant studies \cite{icarus-361-114331,jgr-112-e100}.

\begin{figure}[!hbt]
  \centering
  \includegraphics[width=8cm]{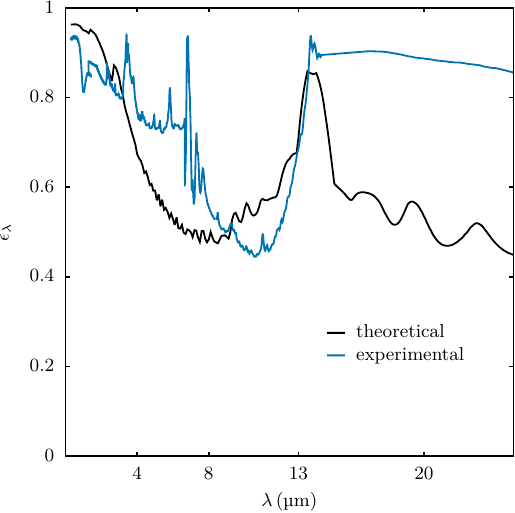}
  \caption{Measured and calculated 
           emissivity/absorptivity of \SI{50}{\micro\metre}-thick 
           ilmenite-pigmented polyethylene film on aluminium substrate
           by employing the optical constants determined using 
           $s=10^{-7}$ \si{\per\micro\metre} in the Hapke model.}
  \label{fig-con-s-ems}
\end{figure}

Employing the determined optical constants using small internal 
scattering coefficient of $s=10^{-7}\,\si{\per\micro\metre}$, the 
emissivity/absorptivity of the \SI{50}{\micro\metre}-thick ilmenite-pigmented 
film on aluminium is calculated and shown in Fig.~\ref{fig-con-s-ems}.
Comparing with the measurement data, it is observed that a good 
agreement in the spectral range of $\lambda<\SI{13}{\micro\metre}$ 
is achieved, although acceptable discrepancy exists.
However, an overwhelmingly underestimation in emissivity and  
absorptivity is found in the spectral range of $>\SI{13}{\micro\metre}$.
This phenomenon agrees with the characteristics of the standard Hapke 
radiation model, which is known to be incapable of addressing the strong 
absorption features of absorbing materials.
\subsection{Wavelength-dependent internal scattering coefficient}
\begin{figure}[!hbt]
  \centering
  \includegraphics[width=8cm]{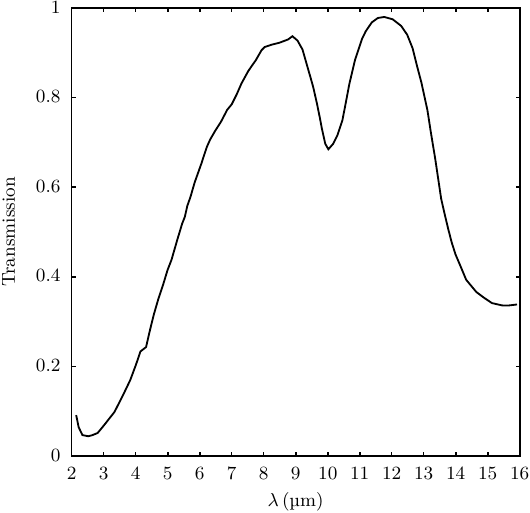}
  \caption{Transmission spectra of ilmenite powder deposited on
           a conventional rock salt window.}
  \label{fig-inf-tra}
\end{figure}

\begin{figure}[!hbt]
  \centering
  \includegraphics[width=8cm]{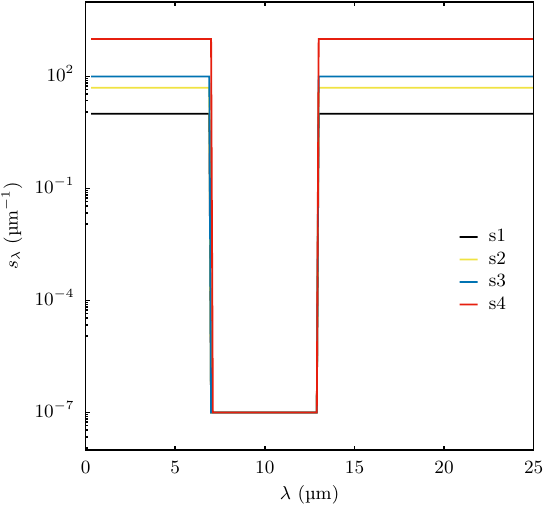}
  \caption{Wide band internal scattering coefficient.}
  \label{fig-vary-s}
\end{figure}

\begin{figure}[!hbt]
  \centering
  \includegraphics[width=8cm]{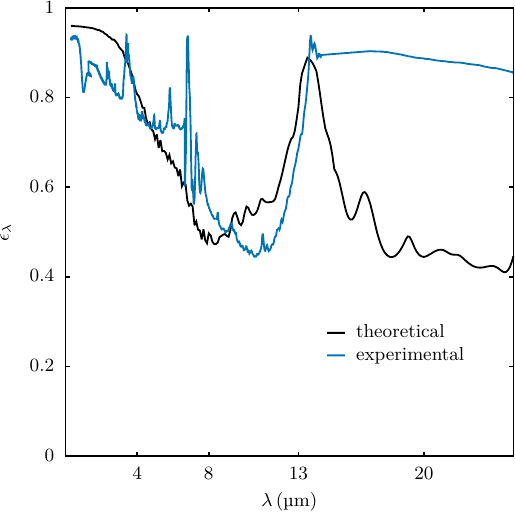}
  \caption{Measured and calculated 
           emissivity/absorptivity of \SI{50}{\micro\metre} 
           ilmenite-pigmented film on aluminium substrate
           by employing the optical constants determined using
           wide band internal scattering coefficient of $s_{1}=10$
           \si{\per\micro\metre} in regions of strong absorption.}
  \label{fig-vary-s1-ems}
\end{figure}

\begin{figure}[!hbt]
  \centering
  \includegraphics[width=8cm]{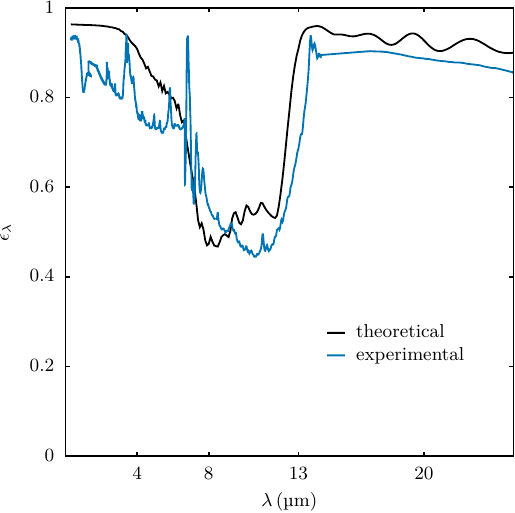}
  \caption{Measured and calculated
           emissivity/absorptivity of \SI{50}{\micro\metre} 
           ilmenite-pigmented film on aluminium substrate
           by employing the optical constants determined using
           wide band internal scattering coefficient of $s_{2}=50$
           \si{\per\micro\metre} in regions of strong absorption.}
  \label{fig-vss1-ems}
\end{figure}

To resolve the issue of underestimation in the infrared absorption 
recognized above, assumption of constant value of the internal 
scattering coefficient across the entire spectral range, 
which is usually employed for unknown materials 
in the standard approach, is required to be released.
Fortunately, more supporting information on absorption features 
of ilmenite material is available in a relevant study\cite{ac-022-1478}. 
Ilmenite powder having an average particle diameter smaller 
than \SI{5}{\micro\metre} was deposited as a film on a conventional 
rock salt window for infrared absorption investigation and
the transmission spectra are shown in Fig.\ref{fig-inf-tra}.
The curves have shown high transmission in the spectral 
range of 7--\SI{13}{\micro\metre}, and are observed to absorb 
strongly over spectral range starting at \SI{13}{\micro\metre}.
Thus, in the current study, only wide band specifications 
($\lambda<\SI{7}{\micro\metre}$, 
$\SI{7}{\micro\metre}<\lambda<\SI{13}{\micro\metre}$,
$\lambda>\SI{13}{\micro\metre}$) are taken into account 
for revealing distinct, wavelength-dependent absorbing features 
considering that much better resolution spectral data are not available.
The internal scattering coefficient in regions of weak absorption
require a low value to generate the phenomenon associated with
the measured reflectance data.
In the regions of stronger absorption, the trend associated
with the reflectance spectra require larger values of the 
internal scattering coefficient.
Thus, modification of constant values of $s$ is made by specifying large 
values on the internal scattering coefficient in the spectral range of 
strong absorption when the wavelength is smaller than 
\SI{7}{\micro\metre} and larger than \SI{13}{\micro\metre}. 
As discussed and concluded in the previous subsection, 
$s>10^{-1}\,\si{\per\micro\metre}$ is found to be the 
delineation between small and large values of $s$. 
Thus, in the spectral range of strong absorption 
($\lambda<\SI{7}{\micro\metre}$ and $\lambda>\SI{13}{\micro\metre}$), 
values of $s=10, 50, 100, 1000$ \si{\per\micro\metre} 
are employed in the calculation for determining the optical properties.
The wavelength-dependent internal scattering coefficient 
is then plotted in Fig.~\ref{fig-vary-s}.
The individual optical determination by employing the varied 
internal scattering coefficient is subsequently conducted.
Then, the emissivity/absorptivity of the film system using
the determined optical properties is calculated. 
The curve shown in Fig.~\ref{fig-vary-s1-ems} is obtained by employing 
$s_{1}=10$ \si{\per\micro\metre} in regions of strong absorption.
It is clear that the underestimation of absorption in the 
infrared region remains. 
Increasing the value to $s_{2}=50$ \si{\per\micro\metre}, this 
issue is resolved and accurate prediction of emissivity/absorptivity 
in the region is achieved (see Fig.~\ref{fig-vss1-ems}).
Additionally, it is found that reconstructed reflectance spectra 
agree well with those of measurements.
A continuing increase in the internal scattering coefficient ($s_{3}, s_{4}$)
in regions of strong absorption gradually fails both the absorption
modification in the infrared regions and the fitting agreement 
in reflectance spectra.
This thus shows accuracy and validity for determination of the internal
scattering coefficient and optical properties.
\subsection{Optical properties}
\begin{figure}[!hbt]
  \centering
  \includegraphics[width=8cm]{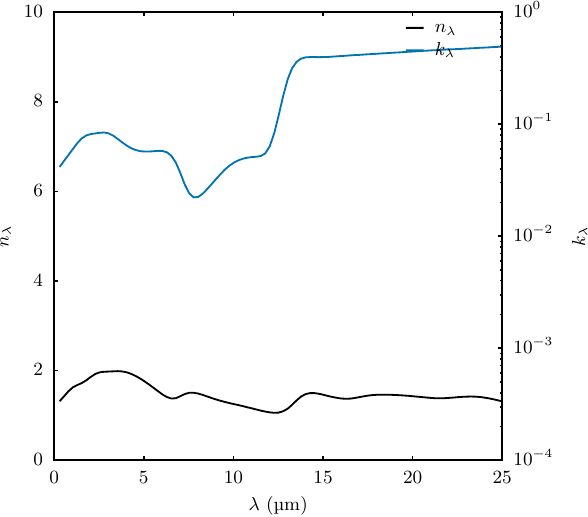}
  \caption{Optical constants of ilmenite material.}
  \label{fig-vss1-nk}
\end{figure}

Employing the wavelength-dependent internal scattering coefficient of
$s_{2}=50$ and $10^{-7}$ \si{\per\micro\metre}
in regions of strong absorption and high transmission, respectively,
the optical properties of ilmenite material are thus
obtained and shown in Fig.~\ref{fig-vss1-nk}.
The value of the refractive index $n$ varies weakly on the wavelength.
In contrast, a more pronounced change in the absorptive index 
$k$ with the wavelength is observed. 
Specifically, low values of the absorptive index $k$ on the 
magnitude of $10^{-2}$ in the atmospheric window spectral range 
are observed, which results in high transmission 
of ilmenite in the spectral range.
Furthermore, in the strong absorption infrared spectral range starting 
from \SI{13}{\micro\metre}, values of the absorptive index $k$ are higher 
than 0.1, which is also in agreement with the statement of $k>0.1$ 
for absorbing ilmenite demonstrated in the study \cite{jgr-086-3039}.
\section{Conclusions}
We recognize the necessity of determination for optical properties of
absorbing ilmenite and issues associated with bi-directional
reflectance spectroscopy identification method used in this work.
The spectroscopy consists of measurements on reflectance and analysis
of radiative transfer in particulate media by multiple 
scattering theory (Hapke radiative transfer model).
One issue encountered in this standard approach is the handling of
the refractive index $n$, which is assuming a constant 
value subject to a prior understanding of the materials, 
and subsequently employing a subtractive Kramers--Kronig 
analysis to allow for wavelength-dependent values. 
However, this results in fundamental insensitivity to spectral features 
and changes in the absorptive index $k$ across the entire spectral range.
Another issue is that the standard approach provides a poor 
determination of the optical properties in the infrared regions 
where absorbing materials exhibit strong absorption specifications.

In this work, we resolve these issues by complementing distinct 
radiative transfer analysis and then identify the optical properties 
of ilmenite material.
We supplement the Hapke model with an additional analysis for the 
problem of radiative transfer in a slab medium of the material to solve 
the issue of two unknowns $n$ and $k$ and fundamental insensitivities
associated with the subtractive Kramers--Kronig analysis.
The two-flux approximation is used in solving and analyzing 
the radiative transfer within the slab medium.
Aiding with the transmission spectra of the particulate thin film,
the internal scattering coefficient parameter of grains
which is neglected in the standard approach is found to be
the key in identifying the absorption features of materials.
The validation of the combined spectroscopy proposed in this work
and the determination of the internal scattering coefficient
are achieved by scrutinizing the absorptivity/emissivity of the 
multi-layer medium of the material, which is modelled as 
a $N$-phase stratified medium and analyzed by investigating 
the propagation of electromagnetic radiation within.

The results show that a constant internal scattering coefficient, 
which is employed in the standard bi-directional reflectance spectroscopy, 
results in a poor determination of the optical properties in infrared 
regions where absorbing materials exhibit strong absorption specifications. 
Observing the absorption and transmittance features 
from the transmission spectrum of the ilmenite power film,
a wavelength-dependent internal scattering coefficient of 
$s=50$ and $10^{-7}$ \si{\per\micro\metre} in regions of strong 
absorption and high transmission, respectively, is obtained.
The value of the refractive index $n$ varies weakly on the wavelength.
While, a more pronounced change in the absorptive index
$k$ with the wavelength is observed.
Specifically, the low values of the absorptive index $k$ 
on the magnitude of $10^{-2}$ in the atmospheric window 
spectral range corroborate the phenomenon of high transmission 
of ilmenite in this spectral range.
In the strong absorption infrared spectral range starting from
\SI{13}{\micro\metre}, the values of the absorptive index $k$ are 
higher than 0.1, which corroborate strong absorption of the material.
%
%
\footnotesize
\setlength{\bibsep}{0.2cm}
\biboptions{comma,square,sort&compress}
\bibliographystyle{plain}
\bibliography{ms}
\clearpage
\newpage
%
\end{document}